# In situ AFM Observations of Li-Oxygen Electrochemical Reactions

Hariharan Katharajan,[1,2] Kumar Virwani,[1*] A. David Erpelding,[1] and Jeannette M. Garcia[1*]

[1]IBM Almaden Research Center, 650 Harry Road, San Jose, CA 95120

[2]Politecnico di Torino, Department of Electronics Engineering, Italy

**Corresponding Author**

*jmgarcia@us.ibm.com (JMG); kvirwan@us.ibm.com (KV)

*Abstract:*

The morphologies of crystalline lithium peroxide ($Li_2O_2$) discharge products in Li-$O_2$ batteries have been shown to exhibit a dependency on subtle variations within the battery cell-operating environment including exposure to ambient air, moisture, or additives. As a result, imaging battery discharge products in real time under carefully controlled environmental conditions is a challenging obstacle for complete mechanistic understanding of $Li_2O_2$ growth and deposition during discharge in metal-air batteries. Here, we report the design of a completely enclosed cell for high-resolution *in situ* AFM imaging of Li-$O_2$ battery discharge products. The air-and moisture-free electrochemical cell environment enabled the observation of different product morphologies during AFM imaging when LiTFSI in oxygen-saturated tetraethylene glycol dimethyl ether (tetraglyme) solvent was employed. This *in situ* AFM cell development brings complimentary information to various proposed mechanisms for lithium oxygen reaction.

*Main body of the paper*

With the rise of alternative energy technologies, its storage using batteries has taken a center stage in current scientific and engineering endeavors. One of the proposed technologies relies on the energetic combination of one of the most electropositive and lightweight cations (Li$^+$) with one of the most electronegative elements (O$_2$) to form a Li-O$_2$ battery (e.g. Eq 1).[1]

$$2 \text{ Li}^+ + \text{O}_2 + 2 \text{ e}^- \longleftrightarrow \text{Li}_2\text{O}_2 \qquad \text{(Equation 1)}$$

In order to overcome limitations in advancing battery technologies beyond that which can be achieved with lithium-ion, significant attention has been given to the nonaqueous lithium-air battery, which has the highest theoretical energy density of the proposed next-generation battery chemistries.[2, 3] The Li-O$_2$ battery consists of a carbon cathode[4,5] and lithium metal anode and is discharged under a blanket of oxygen atmosphere to form the thermodynamic and reversible discharge product, lithium peroxide (Li$_2$O$_2$). However, many practical limitations to obtaining theoretical energy densities remain due to the insulating nature of Li$_2$O$_2$ as well as insoluble deposits resulting from electrolyte decomposition.[6, 7, 8, 9, 10] Although efforts towards enhancing the solubility of Li$_2$O$_2$ to enhance reversibility of the battery have been made through alternative electrolyte combinations and additives,[11, 12, 13] the principal challenge remains, namely: efficient decomposition of the electrolyte-insoluble Li$_2$O$_2$ solid to gaseous oxygen (O$_2$) and soluble lithium ions. Post-operation analysis on crystalline Li$_2$O$_2$ such as X-ray diffraction (XRD), scanning electron microscopy (SEM) and atomic force microscopy (AFM) imaging have been carried out to study the production of $Li_2O_2$. However, these *ex-situ* methods do not provide insight to growth mechanism of $Li_2O_2$ nor its behavior upon deposition on various cathode surfaces.[14] Therefore, we focused our efforts to design a closed electrochemical cell and image the cathode with AFM as it discharges in the cell. This paper reports on the development of an *in situ* cell for direct observations of lithium-oxygen reaction products using an AFM for the first time.

Figure 1 (a) shows SolidWorks® isometric view and Figure 1 (b) shows a cross-section of the cell. The basic cell consisted of bottom stainless steel in contact with a polished glassy carbon cathode, poly(dimethylsiloxane) (PDMS) separator and annular lithium anode in contact with top stainless steel cap. The cell body was constructed from PEEK® polymer. Both cathode and anode contacts were made with shielded coaxial cables. Great care was taken in the design of the cell to enable its assembly inside an argon-filled glove box with minimum operations. The cell design allowed for centering of the cathode, separator and the annular lithium by the use of a stainless steel cap and two metal posts at the bottom current collector as illustrated in Figure 1. An annular design of the lithium anode was employed to ensure that electric fields generated during the electrochemical reaction between lithium and oxygen were concentric and did not develop a point charge as in a previous studies[15, 16]. The AFM tip holder was made of PEEK® which is inert to electrolyte, salt and electrochemical reaction products within the timeframe of the imaging. All components were cleaned and vacuum dried at 70 °C and the AFM probe was mounted onto the tip holder prior to use. The assembly of glassy carbon cathode, PDMS separator, lithium anode, oxygen saturated electrolyte and the AFM probe holder was completed in an

Argon-filled glove box. The cell was sealed using Viton® bellows stretching from the AFM tip holder to the cell. The cell was then removed from the glove box and mounted onto the AFM for measurements while the electrochemical (EC) reaction occurred between lithium and oxygen. The AFM itself operated in a closed hood with a positive pressure of dry nitrogen gas (although this did not come into contact with the sealed AFM sample under Ar). Silicon probes with silicon-nitride cantilevers were used for tapping mode scanning of the cathode surface, in contrast with previous report[15, 16] where the measurements were performed in contact mode. The electrical connections to the cell were designed to enable cell operation from DC to multiple hundred kilohertz.

Before performing AFM scanning studies and real-time observations it was important to understand the electrochemical discharge characteristics of the AFM cell and establish equivalence to the commonly used Swagelok-type battery. An accepted method to characterize the battery is to measure the cell potential while it supplies a constant discharge current referred to as chronopotentiometry (CP). Figure 2 (red curve) shows a typical discharge curve from the AFM cell, performed in an $O_2$-saturated electrolyte consisting of 150 µL of 1 M LiTFSI in dimethoxyethane (DME) solvent. The electrolyte was purified to have < 100 ppm of water after oxygen saturation. Figure 2 (blue curve) shows discharge curve from a Swagelok type cell with the same electrolyte for comparison. For both cells, a constant discharge current of 500 nA was maintained. Complete discharge point of the AFM cell (i.e. cell death) was defined when the potential dropped below 1 V. The initial open circuit voltage of the AFM cell was 2.90 V. The discharge potential was very close to the ideal open circuit voltage of ~2.91 V expected from such a cell configuration and compares very well with the Swagelock type cell open circuit voltage. As soon as a resistive load is connected to the cell to maintain a constant 500 nA current, the potential drops to 2.6 V in the case of AFM cell identical to the Swagelock type cell. Beyond that point, the AFM cell continued to operate at a constant potential until sudden death, 48 hours later. The total capacity of the AFM cell was ~30 µA-h (Figure 2). During the cell discharge, solids formed on the carbon cathode. Comparatively, the AFM cell discharged completely while the Swagelock type cell fitted with capillaries connected to a positive pressure of $O_2$ continuously supplies a constant potential even after 48 hours, likely due to the diminishment of available oxygen in the saturated solution over discharge. This suggests that the diminished discharge capacity observed in the electrochemical cell was likely due to complete oxygen consumption during discharge; whereas in the Swagelock cell, the cell is kept under a constant pressure of oxygen so cell death was not observed within 48 hours under otherwise identical conditions. The AFM cell discharge measurements were repeated multiple times and helped establish consistency and good equivalency with Swagelock type cell.

Having established consistent discharge characteristics with the AFM cell, *in situ* AFM scanning studies were first performed with 1 M LiTFSI in DME. The cell was mounted on an Asylum Research MFP-3D AFM system. However, the wetting characteristics of the DME electrolyte were such that a good meniscus was not established with the AFM probe holder that resulted in poor AFM imaging. Tetraglyme was thus employed since its boiling point is ~160 °C, twice that of DME and its viscosity is five times greater and has shown to be an effective ethereal solvent for Li-oxygen chemistry.[17, 18] Both these factors allowed formation of a very stable meniscus between the AFM probe holder and the glassy carbon surface. Figure 3 shows images of the same glassy carbon region before and after cell discharge performed *in situ* AFM. The discharge current was gradually ramped up from 500 nA to a final current of 6 µA in the first ten minutes allowing the measurements to complete in about six hours (instead of >48 hours). The drift rate was measured to be <15 nm/min this allowed the same region of the cathode surface to be maintained in the scan view without any difficulty. A comparison of before and after images suggests that the growth of discharge product, $Li_2O_2$ was likely conformal to the underlying glassy carbon. Thus the eventual result of the discharge process is a small change in roughness of the glass carbon from 4 nm to 5 nm. However the change in surface roughness may also be attributed to a change in the AFM tip shape as a function of time. As such the initial *in situ* AFM experiments were inconclusive about the reaction dynamics.

Recent studies from our laboratories and others[19,20] have reported that the introduction of small amounts of water into an ethereal solvent results in the formation of varied crystal shapes of $Li_2O_2$ on the cathode surface based upon SEM observations of the cathodes at the completion of the discharge reaction. In order to compare anhydrous conditions, we studied the effects on the products when 4000 ppm of water was introduced into the electrolyte. We were motivated by the ex-situ observed differences in $Li_2O_2$ morphology after cell operation (i.e. toroid formation vs. conformal coating) and the accompanying reaction dynamics during the electrochemical discharge. Figure 4 (and a movie in supplementary materials) shows a series of *in situ* AFM images collected over a course of six hours of measurements and the corresponding discharge curve when water was present. The AFM was set to collect 256 X 256 pixel resolution images at a scan velocity of 10 μm/sec in tapping mode imaging with silicon probes on silicon nitride cantilevers. Similar to Figure 3 the current was ramped from 500 nA to 6 μA to allow the completion of the study in about six hours. The discharge curve is labeled with letters 'a' through 'g' corresponding to the AFM images labeled 'a' through 'g.' Topography image Figure 4(a) was collected at the start of the electrochemical reaction and Figure 4(b) was collected one hour after the reaction start. There was no measurable difference in the surface topography during the first hour of the discharge. After almost 90 minutes from the start of the discharge lithium-peroxide particles in the size range between 100 nm to about 350 nm appear to precipitate on the glassy carbon electrode surface. Figures 4(c) through 4(f) show that the precipitate[21, 16] formed increased to almost 1 μm diameter and new particles continue to precipitate on the cathode surface. Topography imaging of these hundreds-of-nanometers sized particles resulted in a loss of tip sharpness and invariably its shear and breakage so that even though the discharge reaction continued the AFM imaging was compromised after this point. Clear differences were observed in both the size and morphologies of the solids produced in each environment, and future studies to further characterize these products are underway in our laboratories and will be reported in due course.

Aforementioned observations show for the first time direct *in situ* topography imaging of lithium oxygen electrochemical reaction with a custom designed electrochemical cell. We have resolved challenges associated with assembly and sealing of an AFM cell inside an Ar-filled glove box with the probe mounted onto the probe holder, followed by *in situ* scanning in nitrogen ambient environment. Materials for the cell were chosen to be inert to all the reactants and products of the lithium oxygen electrochemical reaction during the imaging timeframe and challenges arising from solvent de-wetting were successfully overcome. The cell was thermally stable enabling observations of the same region of the cathode surface for about eight hours. In addition, low discharge current (~500 nA) studies showed that the developed *in situ* AFM cell is similar to the commonly used Swagelock type battery in its electrochemical characteristics and stable for over two days. Distinctly different topographical changes on the glassy carbon cathodes were measured for tetraglyme solvent. Although the focus of this study was to achieve stable and consistent *in situ* AFM cell design and scanning, our experiments indicated that it might be possible that the reaction between lithium and oxygen is initiated *in solution* prior to precipitation, evidenced by particles that precipitated from the solution onto the carbon cathode that then grew in size. The use of *in situ* AFM cell for direct observations should enable a more comprehensive understanding of the oxygen reduction reaction and provide experimental evidence to conclusively establish the reaction kinetics and mechanism of $Li_2O_2$ production. These reaction kinetics and mechanisms may vary for different solvents hence an initial representative study is presented in this manuscript. With the development of emerging multimodal imaging techniques for AFMs it is envisioned to use complimentary techniques such as nanomechanical measurements to further the understanding of lithium-oxygen reaction. Future directions will be aimed at the characterization of particles formed in the presence of additives and at other current rates (i.e. >100 μA). Finally, similar studies will be extended to other reactions that could benefit direct observations of electrochemical phenomena.

*Acknowledgements*

The authors are grateful for productive and helpful discussions with Dr. Spike Narayan. H. K. wishes to acknowledge his joint internship program through IBM and the Politecnico di Torino.

*Figures*

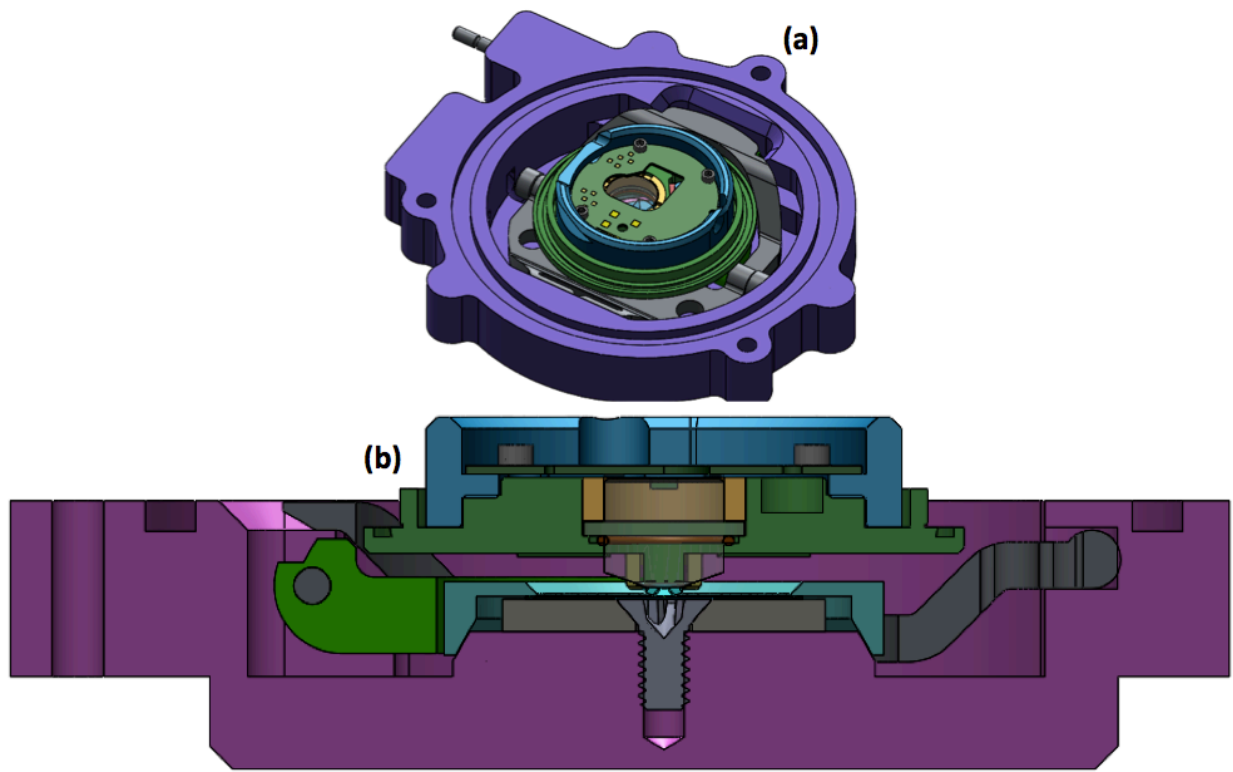

Figure 1: SolidWorks® model of the cell design. (a) Isometric view of the in situ electrochemical cell and the probe holder. (b) Cross-section view through the center of the cell. The image is drawn to scale to illustrate the challenges in design and eventual machining of the cell. Great care was taken to enable assembly of the cell inside a glove box.

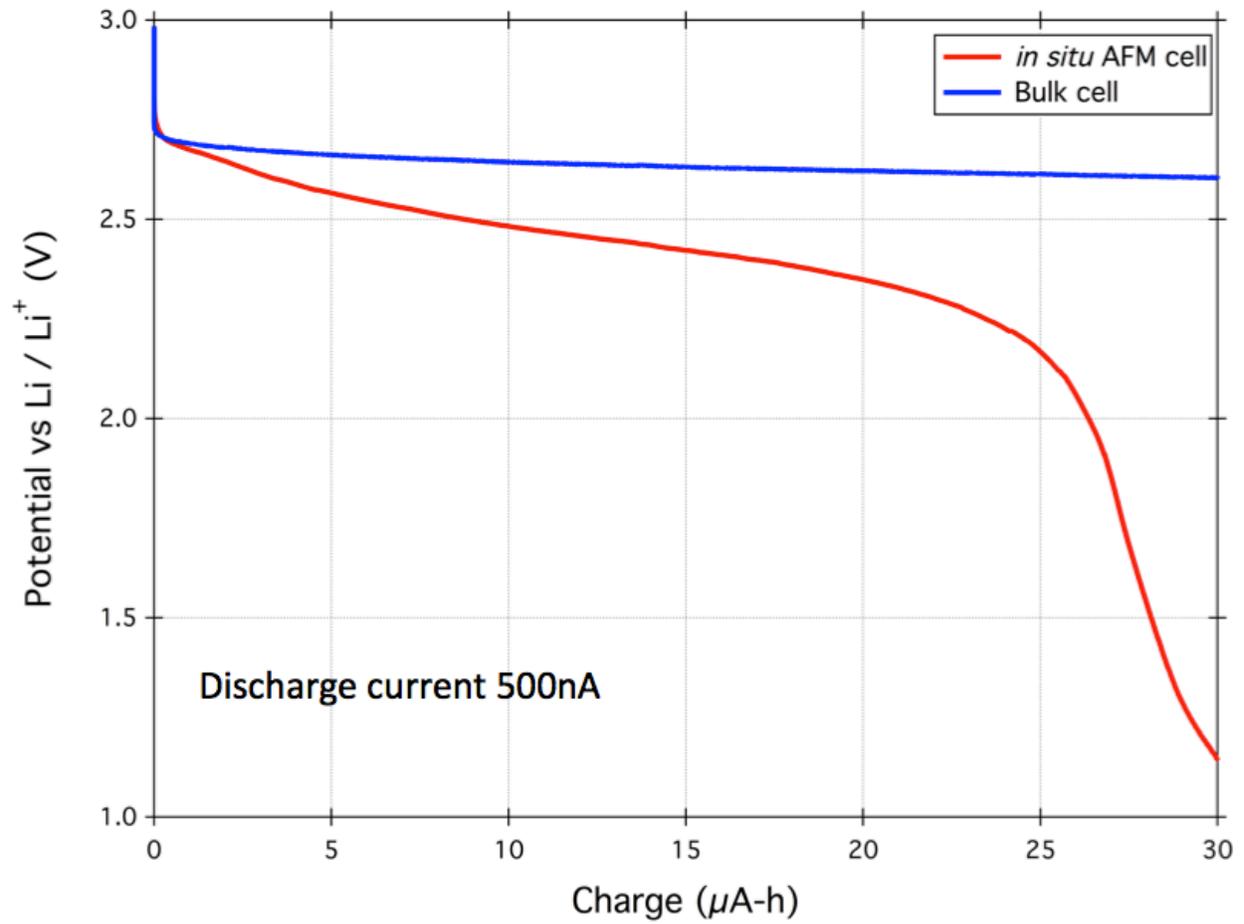

Figure 2: Typical discharge curves from the AFM cell (red curve) and from a Swagelock cell (blue curve) performed in an $O_2$-saturated electrolyte consisting of 150 µL of 1 M LiTFSI in dimethoxyethane (DME) solvent.

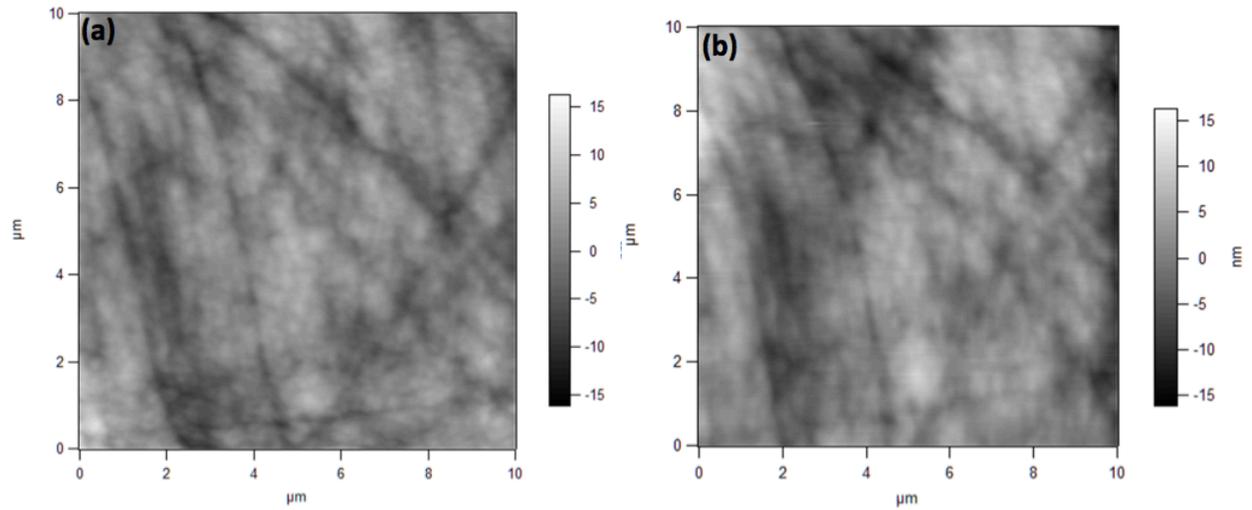

Figure 3: Topography images of the same glassy carbon region before (a) and after (b) cell discharge performed *in situ* AFM with tetraglyme solvent and 1M LITFSI as the electrolyte with < 100ppm of water. There was a 20% change in roughness from 4nm to 5nm after discharge.

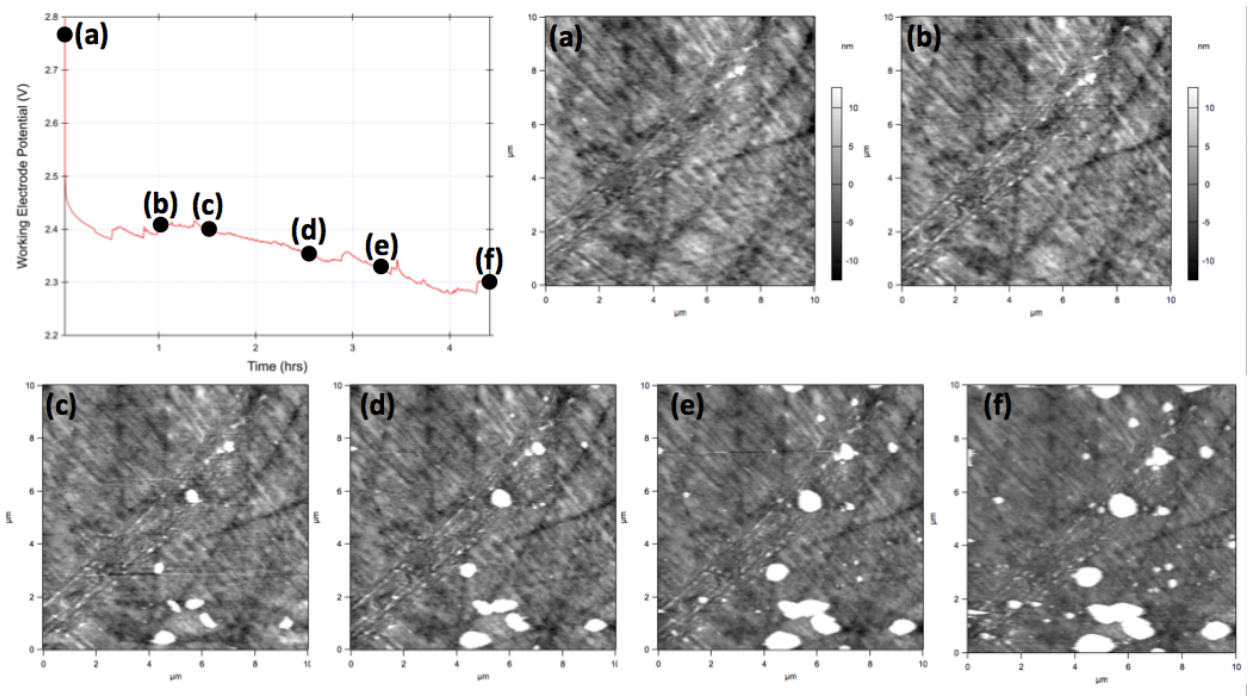

Figure 4: Cell discharge curve along with corresponding *in situ* AFM topography images collected in tapping mode at various time snap-shots. Tetraglyme was used as a solvent and 1M LITFSI as the electrolyte with 4000ppm of water. (a) Surface before the start of the cell discharge. (b) – (f) Topography of the exact same region at 1hr, 1.5hrs, 2.5hrs 3.25hrs and 4.4hrs after the start of the discharge reaction.